\documentclass[aps,pre,twocolumn,groupedaddress, showkeys, a4paper]{revtex4}
\usepackage{amsmath}
\usepackage{graphicx}
\usepackage{hyperref}

\begin{document}

%\preprint{}

\title{Closed benchmarks for network community structure characterization}

\author{Rodrigo Aldecoa}
\email[]{raldecoa@ibv.csic.es}
\author{Ignacio Mar\'in}
\email[]{imarin@ibv.csic.es}
%\thanks{}
\affiliation{Instituto de Biomedicina de Valencia.
Consejo Superior de Investigaciones Científicas (IBV-CSIC)
Calle Jaime Roig 11. Valencia, Spain}

\date{\today}

\begin{abstract}
Characterizing the community structure of complex networks is a key challenge in many scientific fields. Very diverse algorithms and methods have been proposed to this end, many working reasonably well in specific situations. However, no consensus has emerged on which of these methods is the best to use in practice. In part, this is due to the fact that testing their performance requires the generation of a comprehensive, standard set of synthetic benchmarks, a goal not yet fully achieved. Here, we present a type of benchmark that we call "closed", in which an initial network of known community structure is progressively converted into a second network whose communities are also known. This approach differs from all previously published ones, in which networks evolve toward randomness. The use of this type of benchmark allows us to monitor the transformation of the community structure of a network. Moreover, we can predict the optimal behavior of the variation of information, a measure of the quality of the partitions obtained, at any moment of the process. This enables us in many cases to determine the best partition among those suggested by different algorithms. Also, since any network can be used as a starting point, extensive studies and comparisons can be performed using a heterogeneous set of structures, including random ones. These properties make our benchmarks a general standard for comparing community detection algorithms.
\end{abstract}


\maketitle

\section{INTRODUCTION}
Network analysis offers a powerful approach to solve problems in many scientific fields, including physics, biology, and sociology \cite{1,2,3,4}. Community structure is a significant property of these networks. A community can be loosely defined as a set of nodes that are more densely connected among themselves than with the rest of the network. The importance of community structure characterization derives from the fact that all nodes in a community are expected to share common attributes, features, or functional connections (reviewed in \cite{5}). Many algorithms and methods have been proposed for extracting the optimal partition of a network into communities. While some of them try to improve a global quality function such as its Modularity \cite{6} or Surprise \cite{7}, others search for the optimal partition by minimizing the compression of the information that best describes the network \cite{8}, minimizing the Hamiltonian of a Potts-like spin model that represents the graph \cite{9}, or deducing the maximum-likelihood model that best fits the structure of the network \cite{10}, to name just a few examples. However, none of these algorithms achieves maximal results in all situations. Their performance varies greatly, depending on the topological parameters of the
analyzed network \cite{7,11}.

In order to compare the performance of community detection algorithms, several benchmarks have been proposed. The first ones were based on the planted one-partition model \cite{12}. The most popular among them is the Girvan and Newman (GN) benchmark \cite{13}, in which a network of 128 nodes is divided into four communities of equal size where each node is connected with 16 other members of its own community. This starting graph can then be progressively degraded by replacing links within communities with links between them, keeping constant the average node degree. The relaxed caveman (RC) benchmarks \cite{7,14,15} are similar in concept. In them, the starting network is formed by a set of cliques of variable sizes, and a degradation process identical to that already described for the GN benchmark is performed. Notice that GN and RC communities are, by definition, Erd{\H{o}}s-R{\'e}nyi subgraphs \cite{16} in which, all throughout the degradation process, each pair of nodes is linked with the same probability \textit{p}. This makes those benchmarks rather inappropriate for representing real-world networks since the latter exhibit much more heterogeneous degree distributions \cite{17,18}. With this idea in mind, Lancichinetti, Fortunato, and Radicchi developed a novel type of benchmark, called LFR \cite{19}, in which both the sizes of the communities and the distribution of node degrees are adjusted to follow power laws. In LFR benchmarks, the fraction of links $\mu$ that a node shares with nodes in other communities is tunable. Increasing $\mu$ (often called the "mixing parameter") generates an analogous behavior to that of the degradation process described for GN and RC benchmarks, i.e., the proportion of intercommunity links grows and the original communities gradually disappear. We refer to all of these benchmarks (GN, RC, and LFR) as "open", given that the final outcome is "open-ended" (i.e., the precise final community structure of the network is undetermined).

In this paper, we describe in detail a novel type of benchmark (referred to as "closed") that is based on the conversion of a network of known community structure into a second network whose communities are also known. We already introduced the concept of a closed benchmark in a previous work \cite{7}, and we showed how this type of benchmark can be successfully used to compare community detection algorithms. Here, we explain it in detail, give some examples of its performance, and discuss its potential and the significant advantages it presents over the aforementioned open benchmarks. We show that the guided evolution of the networks to a closed end enables us to accurately monitor the transformation progress and to evaluate the goodness of a partition at any moment of the process.

\section{FEATURES OF THE CLOSED BENCHMARKS}

\begin{figure}[ht]
\begin{center}
\includegraphics[scale=1]{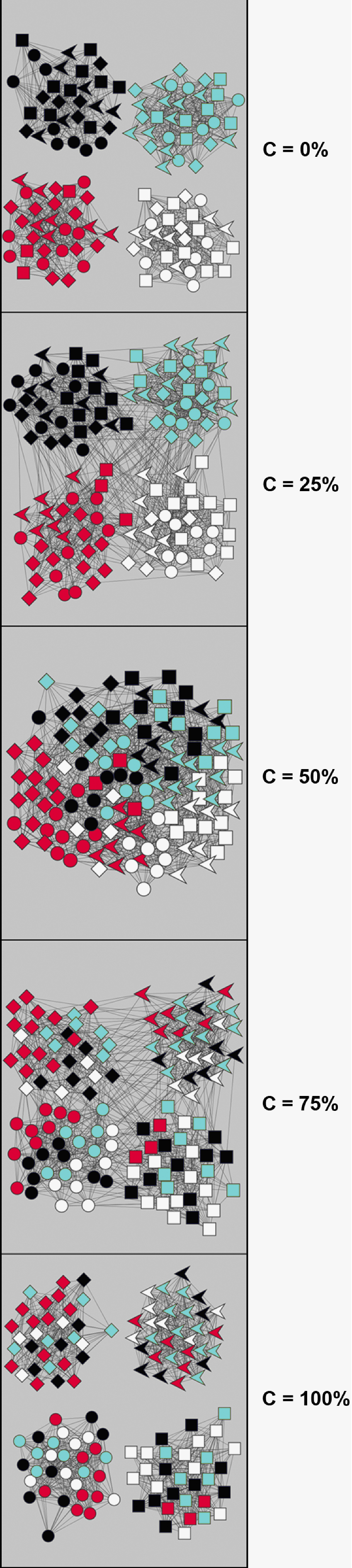}
\caption{\label{fig:1} Transformation process in a closed benchmark. In this case, the starting network is the GN benchmark. Links are progressively rewired from the initial (C = 0\%) to the final network (C = 100\%). Nodes color is defined by the initial community to which they belong, whereas their shape corresponds to the final community in which they are contained.}
\end{center}
\end{figure}

The main concept behind the closed benchmarks is the directed conversion of a network into another one by means of edge rewiring. The starting point is a network whose community structure is known \textit{a priori}. Any type of graph and community structure is valid as an initial network. The algorithm then generates a second, "final" network. The initial and final networks are precisely related. The community structure of the final network is identical to the initial one, but the labels of the nodes are randomly mapped from the former to the latter. Converting the initial network into the final one involves rewiring links in a directed manner, a process depicted in Fig. \ref{fig:1}. The details of the procedure are as follows:
\begin{enumerate}
\item Links present in both the initial and the final networks will not be rewired.
\item At each step, one of the rewirable links is removed and subsequently a new link is added between two nodes connected only in the final network. Conversion ($C$) is defined as the percentage of rewirable links modified at a particular point of the process of converting the initial network structure into the final one.
\item At any point of this conversion process, the network can be saved for later analyses. Therefore, a wide set of intermediate structures to test the behavior of community detection algorithms can be obtained.
\item The process stops when the final structure is reached. 
\end{enumerate}
A significant feature of the closed benchmarks is that, during the conversion process and because of the directed rewiring of the links, we are approaching the final structure at the same rate that we are leaving the initial one. Calling $D$ the distance between both networks, we can assert that the structure at a distance $x$ from the start is also at a distance $D - x$ from the end of the benchmark. This fact, together with the identical topology of both ends, produces a set of structures that is symmetrical about the 50\% conversion point. That is, when $C$ = 50\%, the structure of the network is, on average, at the same distance from both the initial and the final networks. Given these patterns of network evolution, we can assume that its community structure undergoes a similar behavior. As we will describe below, this behavior is central to the evaluation of partitions in closed benchmarks.

Any benchmark is associated with one or several measures of performance. In the case of clustering comparison, several methods, based on counting pairs, cluster matching, or information-theory based indexes, have been developed (reviewed in \cite{5,20}). Among the latter type, the variation of information ($V$) \cite{21} is an information-based distance useful for measuring the dissimilarity between two partitions, $A$ and $B$ ($V_{AB}$). In our context, we consider that it has clear advantages over other criteria, especially its metric nature. This property implies that $V$ is positive-definite, a symmetric distance (which is a highly desirable property when comparing clusterings), and, more important for our purposes, it satisfies the triangle inequality \cite{21}. This last fact turns out to be very useful for closed benchmarks evaluation. In these benchmarks, we have two known community structures, those of the initial ($I$) and final ($F$) networks. Moreover, the method generates a set of intermediate, estimated structures ($E$) whose communities can also be determined. We can deduce from the $V$ triangle inequality the following formula:
\begin{equation}
\label{eq1}
\displaystyle V_{AB} + V_{BC} \geq V_{AC}
\end{equation}

Hence, the sum of $V_{IE}$ and $V_{EF}$ is lower bounded by $V_{IF}$ , which is constant, given that the partitions of the initial and final networks are fixed. If the rewiring of the network has not yet started, the optimal estimated partition is the same as the initial one, $I = E$, and therefore $V_{IE} = 0$ and $V_{EF} = V_{IF}$, satisfying the equality in Eq. \ref{eq1}. When the conversion starts, and because the network approaches the final structure at the same rate that it leaves the initial one, $V_{IE}$ should increase as much as $V_{EF}$ decreases. Therefore, unless the structure of the network becomes very different from both the initial and final structures along the conversion process (e.g., as described in the next paragraph), this should make the equality $V_{IE} + V_{EF} = V_{IF}$ true all along the conversion of the initial into the final structure. A significant deduction is that if, for a given estimated partition $E$, the sum of $V_{IE}$ and $V_{EF}$ deviates from the constant value $V_{IF}$ , then $E$ may not be the optimal partition \cite{7}. Thus, deviation from the expected $V_{IF}$ value may indicate a suboptimal performance of a given algorithm.

If third-party structures, very different from the initial and final ones, are formed along the conversion process, we can find $V_{IE} + V_{EF} > V_{IF}$ even if the partition is optimal. This can be illustrated assuming that the intermediate structure becomes fully random. Two situations are then possible, depending on the density of links in the graph. If, at some point of the rewiring, the intermediate structure becomes a single community containing all the nodes -as expected in a random graph with a high density of links- then $V_{IE} = H(I)$ and $V_{EF} = H(F)$, where $H(I)$ and $H(F)$ are the entropies of the initial and final partitions. Given that $V_{IF} = H(I) + H (F) - 2M(I,F)$, where $M(I,F)$ is the mutual information between the initial and final partitions, we have that $V_{IE} + V_{EF}$ must be somewhat larger than $V_{IF}$. This derives from the fact that $M(I,F) = 0$ only if $I$ and $F$ are independent, which is not the case here. On the other hand, if the density of links is low and the network is randomized, the community structure may approach a situation in which each node is isolated in a different community. If this is true, it can be shown that $V_{IE} = log N - H(I)$ and $V_{EF} = log N - H (F)$, where $N$ is the total number of nodes. In this case, we will find $V_{IE} + V_{EF} \gg V_{IF}$. Thus, if an algorithm is performing perfectly well ($V_{IE} + V_{EF} = V_{IF}$) until a certain conversion percentage, and if, when conversion progresses further, we find $V_{IE} + V_{EF} > V_{IF}$ , this may be due to two reasons: (i) a bad performance of the algorithm with poorly defined community structures, (ii), the emergence of a third-party, potentially random, community structure. This interesting situation will be illustrated in a particular case below.

\section{TESTS}
\subsection{Configuration}
As mentioned above, the particular features of a network can greatly influence the ability of a given algorithm to detect its community structure. For this reason, we performed tests on computer-generated networks that varied in size, node degree distribution, number of communities, and also community sizes. This last parameter has been shown to be crucial in community detection \cite{7,11}. There are two main reasons for the significant effect of community size variation. First, networks presenting a skewed distribution of community sizes are more rapidly degraded than those with equally sized communities because of the quick destruction of small clusters. Second, a skewed distribution may greatly affect the performance of particular algorithms. For example, any algorithm maximizing a popular global measure for community detection, Newman and Girvan's modularity ($Q$), will have trouble detecting small communities, given that $Q$ is affected by a resolution limit \cite{22}.

\begin{figure*}[ht]
\centering\includegraphics[scale=1]{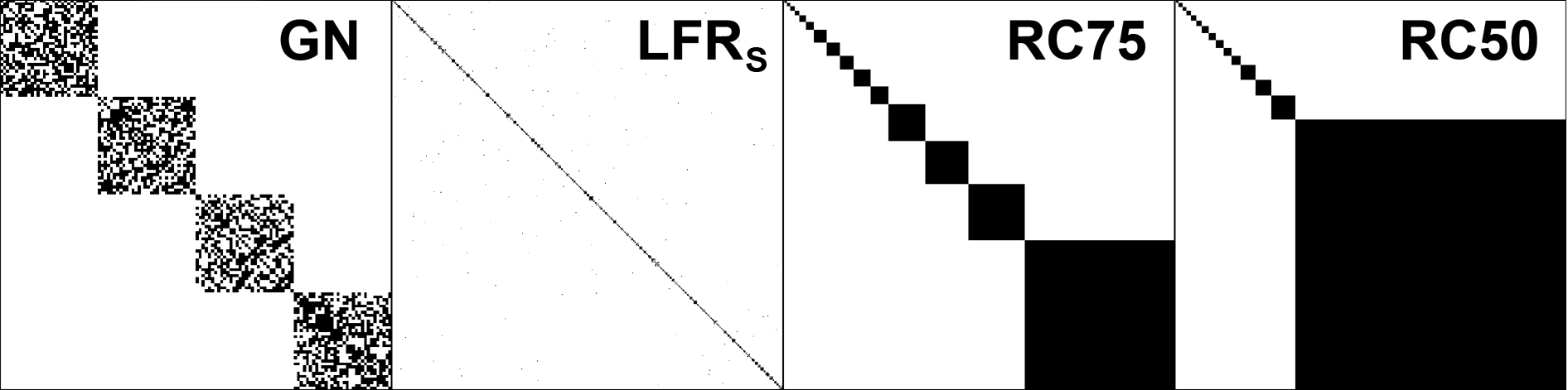}
\caption{\label{fig:2} Graphical view of the adjacency matrices of the four initial networks used in the tests. Nodes are ordered according to the communities to which they belong. Black indicates that two nodes are connected. Differences in relative community sizes are evident. In the GN network, the nodes of the four equal-sized communities are sparsely connected. The groups in the LFR$_S$ are also sparse. However, there are so many of them (195) that visualization is difficult at this resolution level. The RC initial networks (RC75 and RC50) are formed by 16 cliques and the distribution of community sizes is highly skewed, especially in RC50, where a single community dominates the network.}
\end{figure*}

A suitable way to measure and compare the distribution of community sizes is using Pielou's index ($P$), which quantifies how similar are the groups into which a system is divided. This index takes a value of 1 for equal-sized groups and decreases with increasing size variance \cite{23}. In this study, we chose as starting points four different synthetic networks with different P values that correspond to those of already published open benchmarks. We will name them according to the following convention: (i) Girvan-Newman (GN) \cite{13}: already mentioned above. A network of 128 nodes is divided into four communities of equal size ($P = 1$). Nodes are connected only with members of their own community with an average degree of 16. (ii) Lancichinetti-Fortunato-Radicchi network with small communities (LFR$_S$) \cite{11,19}: a network of 5000 nodes. The average degree of the nodes is 20, their maximum degree is 50, the exponent of the degree distribution is -2, and the exponent of the community sizes distribution is -1. The sizes of the communities vary between 10 and 50 nodes (hence the term "small communities"). Among the many networks that can be generated with these parameters, we chose one containing 195 communities of similar sizes ($P = 0.98$). (iii) relaxed caveman \cite{14} with Pielou's index = 0.75 (RC75): Because a more skewed distribution of community sizes was required to analyze the behavior of the algorithms in a wider range of network structures, we generated a network of 512 nodes with $P = 0.75$, which corresponds to a division into 16 communities, each of them including from 2 to 196 nodes. In the RC75 configuration, the initial network consisted of unconnected communities, each one maximally connected internally, i.e., forming a clique. (iv) Relaxed caveman, $P = 0.50$ (RC50): this has an even more extreme variation in community sizes. The initial network is also comprised of 512 nodes forming 16 cliques, but now the largest one contains 354 nodes. Figure \ref{fig:2} graphically displays the pattern of connections of each of these four initial networks. Once obtained, they were progressively modified by increasing $C$, finally obtaining from each one a set of 101 network structures spanning the whole range from $C = 0$ (initial structure present) to $C = 100$ (final structure present). The corresponding open benchmarks, with the same starting community structures and progressive degradation toward randomness, were also analyzed following standard methods described in previous papers (see, e.g., \cite{13,14,19}). We also discuss below in some detail closed benchmarks with random initial structures.

\subsection{Algorithms}
Two community detection algorithms that have shown an excellent performance in recent studies, namely Infomap \cite{8} and SCluster \cite{15}, were used in this work. Infomap understands finding the community structure of a network as an information compression problem, detecting communities while compressing the topology of the network. It has achieved excellent results on the LFR benchmarks \cite{7,11}. On the other hand, SCluster uses a completely different approach. Using iterative hierarchical clustering \cite{15,24}, the algorithm computes the pairwise distances of the nodes from partial clustering solutions. Subsequently, it constructs a hierarchical tree from which the partition of maximum Surprise \cite{7} is chosen as the optimal solution. Surprise is a quality function that estimates the goodness of a partition based on the comparison between the graph and the null model generated by a random distribution of links \cite{7,24}. SCluster has demonstrated an ability to extract high-quality partitions when dealing with networks whose communities strongly vary in size \cite{7,15}. Moreover, as a third way to extract the best clustering of the network, we selected from the Infomap and SCluster solutions the one with the highest surprise, given that we showed before that surprise maximization not only qualitatively outperformed maximizing the most commonly used global index, namely Newman and Girvan's $Q$, but it also improved the solutions generated by any single algorithm \cite{7}.

Figure \ref{fig:3} illustrates the results of the three methods in our four closed benchmarks. Each partition estimated along the conversion process is compared, using the variation of information, with both the initial (black circles) and the final (red squares) community structures. $V = 0$ means that the partitions compared are exactly the same. We previously mentioned how the sum of the variation of information from an estimated point to the initial and to the final optimal partitions ($V_{IE} + V_{EF}$) should optimally be constant and equal to the $V$ between the initial and the final partition ($V_{IF}$). For visualization reasons, half of this sum ($\overline{V} = [V_{IE} + V_{EF}]/2$) is shown in the figures as a dashed line. $\overline{V} = V_{IF}/2$ is expected if the partition is optimal.

\subsection{Results}
\begin{figure*}[ht]
\includegraphics[scale=1.15]{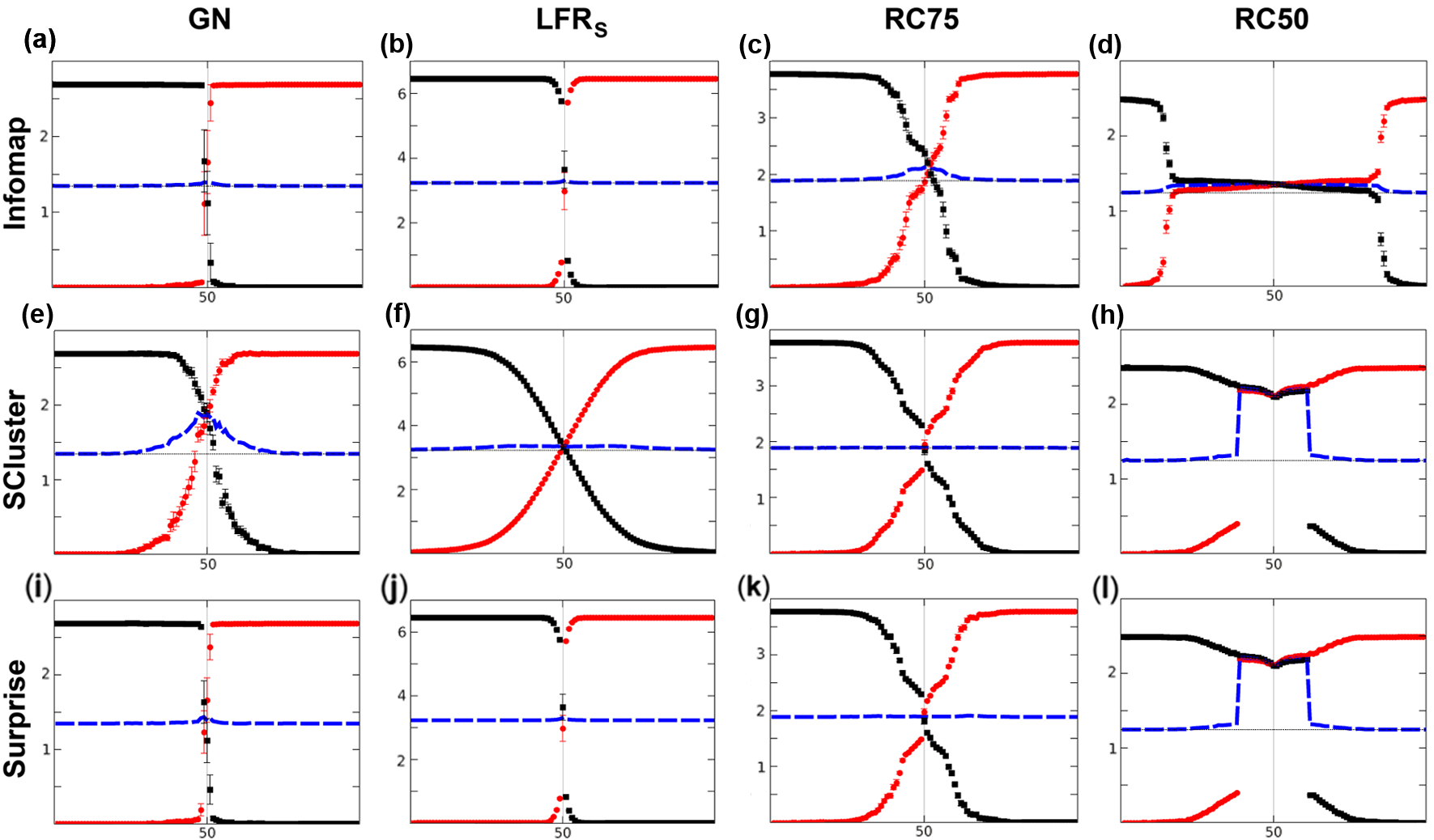}
\caption{\label{fig:3} Variation of information behavior in the four closed benchmarks used in this study. Black circles depict the $V$ between the initial and the estimated partition ($V_{IE}$). Red (gray) squares show the $V$ between the estimated and the final partition ($V_{EF}$). $\overline{V}$ appears as a dashed line, which should follow a straight line if the performance of the algorithm is optimal during the whole process of conversion (i.e., $V_{IE} + V_{EF} = V_{IF}$).}
\end{figure*}

The plots show how different is the community detection process, depending on both the algorithm applied and the topology of the network analyzed. When using the GN network as an input, Infomap performs very well [Fig. \ref{fig:3}(a)]. The variation of information between the initial and the estimated partition ($V_{IE}$, black dots) is zero or near zero along the first half of the benchmark. Moreover, when the conversion ($C$) breaks the 50\% mark, the V between the estimated and the final partition ($V_{EF}$, gray squares) behaves in the same way. That is, the algorithm recognizes the initial structure until $C = 49\%$ and the final one above $C = 51\%$. This is not the case when applying SCluster [Fig. \ref{fig:3}(e)], which only recognizes the initial partition up to $C = 30\%$ and starts recognizing the final partition beyond $C = 70\%$. As expected, $\overline{V}$ graphically shows this different quality in the performance of both algorithms. While in the Infomap plot $\overline{V}$ falls in an almost straight line, matching $V_{IF}/2$, the partitions estimated with SCluster produce a significant deviation from that line in the interval $30-70\%$, where we already detected that the communities were poorly estimated.

\begin{figure*}[ht]
\includegraphics[scale=1]{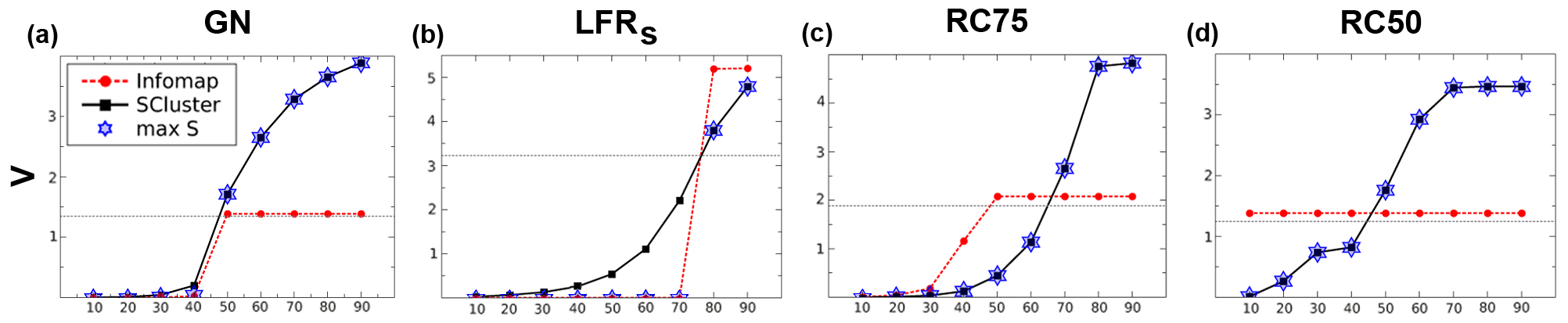}
\caption{\label{fig:4} Open benchmarks with starting structures identical to the initial structures of the closed benchmarks shown in Fig. \ref{fig:3}. These structures are progressively degraded by randomly shuffling links. The percentage of rewired links is indicated on the x axis. The dashed line indicates the $V_{IF}/2$ value of the corresponding closed benchmark. Stars indicate the partitions with the highest surprise values.}
\end{figure*}

When the input of the benchmark is the LFR$_S$ network, Infomap also produces a symmetrical plot, with $\overline{V}$ almost	perfectly matching $V_{IF}/2$ [Fig. \ref{fig:3}(b)]. SCluster also shows in	this case a symmetrical performance, although with a slight	deviation from the optimal values [Fig. \ref{fig:3}(f)], i.e., working	again worse than Infomap. In these first two examples, the sizes of the communities are equal or very similar ($P \approx 1$), and they are expected to be degraded, on average, at the same time. The original partition is thus present during the	first half of the conversion (giving $V_{IE} \approx 0$), and then the	community structure suddenly swaps to the final one (and	then $V_{EF} \approx 0$). On the other hand, when analyzing networks with a strongly skewed distribution of community sizes (RC75, RC50), the performance of the algorithms radically changes. In the RC75 test, Infomap exhibits a nonsymmetrical behavior [Fig. \ref{fig:3}(c)], with $\overline{V} > V_{IF}/2$ when $C = 40-60\%$. On the contrary, SCluster shows a symmetrical pattern with $\overline{V} = V_{IF}/2$ [Fig. \ref{fig:3}(g)]. We can see how the $V$ between the initial and the estimated partition ($V_{IE}$ , black circles) is equal to zero until around the 30\%, at which point it starts to increase. It is very significant that, in an open benchmark (see, e.g., Refs. \cite{13,14,19}), this would be the only available information. Thus we might conclude that from $C = 30\%$ on, these two algorithms fail to recognize the optimal partition. However, a bad algorithm performance is not the only explanation for such patterns. Alternatively, it is possible that the initial partition must not be detected as optimal anymore because the community structure has changed. The closed benchmarks offer a solid way to check if this latter hypothesis is correct. In Figs. \ref{fig:3}(c), \ref{fig:3}(g), and \ref{fig:3}(k), we can see that, although $V_{IE}$ soon starts to grow, $V_{EF}$ begins to decrease at the same rate. That is, the community structure of the initial partition is shifting toward the final one much before the $C = 50\%$ mark is passed, a pattern that is due to the rapid destruction of small communities, typical of benchmarks with low $P$. This behavior was impossible to check in any of the benchmarks published so far, although it is critical for algorithm evaluation. Now, we can assert that the behavior of SCluster is optimal, given that V follows a straight line: it satisfies the equality in Eq. \ref{eq1} during the whole conversion process. In the last case, RC50, the performance of the algorithms follows a pattern that is a bit different from the rest of the benchmarks. Infomap seems to rapidly collapse, with $\overline{V}$ moving away from the optimal straight line, when $C \geq 10-12\%$ [Fig. \ref{fig:3}(d)]. In the case of SCluster [Fig. \ref{fig:3}(h)], $\overline{V}$ values are close to the line quite a bit longer (C around 30\%), but then the algorithm starts recognizing third-party structures, far away from both the initial and final partitions ($\overline{V} > V_{IF}/2$). These behaviors are due to the extremely skewed distribution of community sizes, with a very large group that dominates the network [Fig. \ref{fig:2}(d)]. For these reasons, a quasi-random graph is formed as the conversion process of the benchmark approaches 50\%. Infomap interprets this situation as if most of the network is included into a single community. Hence, as we discussed above, $V$ approximates $H(I)$ (which in this example takes a value of 1.38). SCluster, on the other hand, interprets the network structure as including many singletons. Therefore, V becomes much larger than $\overline{V}_{IF}/2$ for the reasons previously discussed.

Figures \ref{fig:3}(i)-3(l) show the evolution of each benchmark using as the estimated partition that with the highest Surprise between the solutions provided by the two algorithms. As expected \cite{7}, this approach always selects the best partition between those two. The equality in Eq. \ref{eq1} is satisfied all along the first three networks. In the fourth case, the pattern is identical to that produced by SCluster. The Surprise values of the RC50 benchmark suggest that the SCluster interpretation, defining many small clusters of the quasi-random intermediate structure generated when $C > 30\%$, is preferable to the one suggested by Infomap (dominated by a single huge cluster), in good agreement with the fact that SCluster is, as already indicated above, performing better in this benchmark than Infomap in the adjacent conversion range ($30\% \geq C \geq 12\%$).
For comparative purposes, we also generated the corresponding open benchmarks, which start with the same structures as those of our closed benchmarks but are then progressively degraded toward undetermined, random structures by rewiring their links \cite{13,14,19}. Figure \ref{fig:4} shows the variation of information between the original partition and those obtained by the SCluster and Infomap algorithms. The partition with maximum Surprise is marked with a star. We have also depicted in Fig. \ref{fig:4} the value of $\overline{V}$ in the corresponding closed benchmarks (dashed line). As found before in related cases \cite{7}, neither of the two algorithms is the best in all situations. If we use the surprise values as a guide, it can be seen that SCluster improves upon Infomap when degradation is very high and systematically in the benchmarks with the lowest Pielou's indices (RC75 and RC50), while Infomap works better when degradation is low and Pielou's index is high (see GN and LFR$_S$ benchmarks). This situation is fundamentally caused by Infomap solutions often consisting in single communities (this happens in all the cases shown in Fig. \ref{fig:4}, in which the Infomap $V$ values are above the $\overline{V}$ dashed lines). These results for the open benchmarks are fully compatible with those shown in Fig. \ref{fig:3} for closed benchmarks.

\begin{figure*}[ht]
\includegraphics[scale=1]{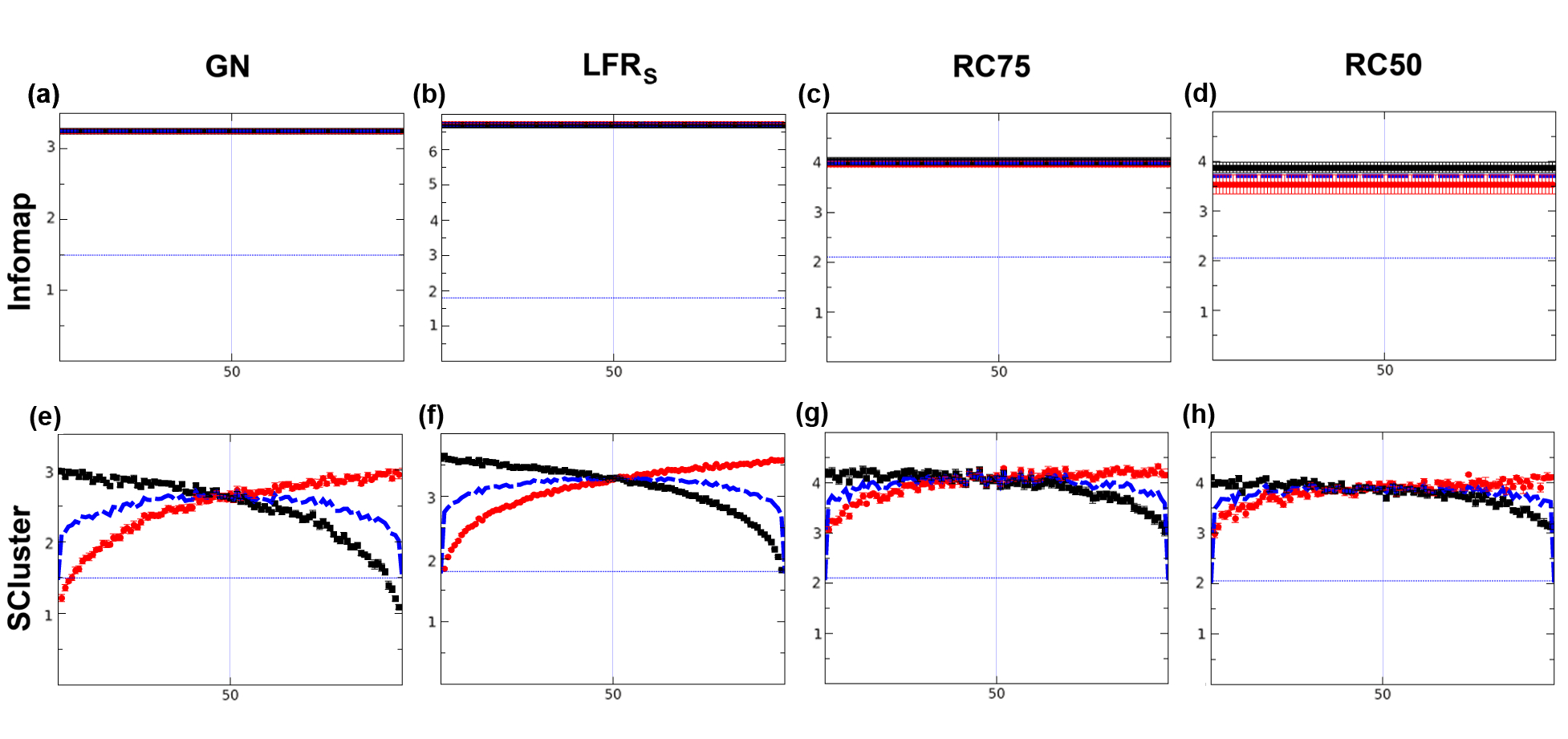}
\caption{\label{fig:5} Random networks with the same number of nodes as the corresponding closed benchmarks indicated on top. As in Fig. \ref{fig:3}, the dashed line corresponds to the $\overline{V}$ value, while red (gray) dots correspond to $V_{EF}$ values and black squares to $V_{IE}$ values. The values of $V_{IE}$, $V_{EF}$, and $\overline{V}$ largely/fully coincide in Infomap analyses, appearing as a single line or close parallel lines. Notice that as soon as conversion starts, $V_{IE} + V_{EF} \gg V_{IF}$. Differences between Infomap and SCluster are due to the different way they interpret the random structures present, i.e., as a single cluster (Infomap) or as many individual clusters (SCluster).} 
\end{figure*}

The comparison of the values of $\overline{V}$ in Figs. \ref{fig:3} and \ref{fig:4} enables us to precisely understand the relationships between both types of benchmarks. Looking at the dashed lines in those figures allows us to estimate the approximate difficulty of reconstructing the community structure present in the closed benchmarks when compared with the open ones. Thus, we can see that $C = 50\%$ in the GN closed benchmark corresponds to a rewiring percentage of more than 40\% in the corresponding open GN benchmark, while $C = 50\%$ in the LFR$_S$, RC75, and RC50 closed benchmarks may correspond, respectively, to rewiring about 80\%, 60\%, and (this can be ascertained less precisely) 50-70\% of the links in the corresponding open benchmark. Thus, the GN, LFR$_S$, and RC75 closed benchmarks always have a substantial level of structure, which explains the good fit to the $\overline{V}$ value observed in Fig. \ref{fig:3}.
Random networks can also be used as starting points for a closed benchmark. The comparison with these random network-based benchmarks may contribute to determine whether or not a given network has a statistically significant community structure, a topic that has recently received some attention \cite{25,26}. To address this issue, we generated four types of random graphs, each of them having the same number of nodes and edges as one of the initial networks described above (GN, LFR$_S$, RC75, and RC50), but randomly distributed. Given that, for generating a closed benchmark, a community structure must be assumed \textit{a priori}, Infomap and SCluster were tested in those random networks and the community structure with the highest surprise value was selected. Figure \ref{fig:5} shows the results of closed benchmarks generated using the four random networks. As occurred above, Infomap returns partitions in which all nodes [Figs. \ref{fig:5}(a)-5(c)], or at least more than 90\% of the nodes [Fig. \ref{fig:5}(d)], belong to one community. The $\overline{V}$ observed is the entropy of the initial (or final) partition $H(I) = H(F)$, given that, if all nodes are in a single community, $H(E) = 0$. On the other hand, SCluster generates solutions with a high number of communities [Figs. \ref{fig:5}(e)-5(h)], interpreting that even a random graph contains a certain degree of community structure. In these random graph benchmarks, an interesting point is to appreciate the extremely fast degradation of the partitions when only 1\% of the links have been rewired (Fig. \ref{fig:5}). When compared with its analogous nonrandom network, $V_{IE}$ rises instantaneously, which is the behavior expected for networks in which communities are barely defined. This kind of comparison between variation of information patterns may enable us to evaluate the robustness of a network, similarly to what has been done using other methods \cite{25}.

\subsection{DISCUSSION}
The development of methods that can accurately detect community structure in networks is critical in many scientific fields, since they can reveal deep underlying relationships among the elements of a system. Therefore, it is very important to compare and evaluate such methods against a set of synthetic benchmarks in order to select one method, or a combination of methods, that can produce reliable results when analyzing real-world networks. Several standard benchmarks for testing community detection algorithms have been proposed, most of them of the class we called open: they start with a network of well-defined community structure and then the structure is degraded by randomly rewiring links \cite{13,14,19}. During this process, the communities gradually disappear toward an "open end" when the precise community structure is undetermined. This type of benchmark is useful for comparing the relative performance of algorithms but inadequate for assessing their intrinsic quality (i.e., whether the solutions provided are optimal or not). In this paper, we have fully described the closed benchmarks, which also degrade an initial network with defined communities, but this time evolving toward a second, known network structure. This evolution is produced by a directed rewiring of the links from the initial to the final network, and it enables us to control the progression of the structure between both ends. We have also shown that the variation of information provides valuable information about the goodness of a partition and its possible optimality: the configuration of our closed benchmarks allows us to lower bound the expected V value using the triangle inequality that the metric must satisfy. Another relevant improvement over the available open benchmarks is the fact that any network can be used as input for the degradation process, enabling us to carry out extensive studies over a wide variety of network topologies. These features clearly represent qualitative improvements over the benchmarks published so far. The comparisons of open and closed benchmarks, or of networks of known structure and random networks (Figs. \ref{fig:4} and \ref{fig:5}), are also interesting ways to further develop this methodology.

As we have shown, there may be scenarios with very skewed distributions of community sizes, such as the RC50 network (Fig. \ref{fig:3}), where the equality in Eq. \ref{eq1} is not satisfied during the whole process of conversion. Nevertheless, this behavior in such extreme networks does not diminish the potential of our approach because, even then, there are several conditions that a good algorithm must fulfill. First, when 50\% of the links have been rewired, $V_{IE}$ must be, on average, equal to $V_{EF}$. Second, the initial partition has to be recognized better than the final one during the first half of the benchmark and, from there on, the behavior should be exactly the opposite. Third, a good algorithm will provide solutions with $V_{IE} + V_{EF} = V_{IF}$ along a longer range of the conversion process than a bad one. In summary, the properties of the closed benchmarks make them highly valuable for the development and evaluation of computational methods to effectively characterize the community structure of a network.

\bibliography{CB}

\end{document}